\documentclass[10pt,twocolumn,letterpaper]{article}

\usepackage{cvpr}
\usepackage{times}
\usepackage{epsfig}
\usepackage{graphicx}
\usepackage{amsmath}
\usepackage{amssymb}

\usepackage{amsmath,amsfonts,bm}

\def\eqref#1{equation~\ref{#1}}

\def\1{\bm{1}}

\def\rmC{{\mathbf{C}}}
\def\rmD{{\mathbf{D}}}

\def\rmK{{\mathbf{K}}}

\def\rmM{{\mathbf{M}}}

\def\rmP{{\mathbf{P}}}

\def\rmT{{\mathbf{T}}}

\def\rmX{{\mathbf{X}}}

\def\va{{\bm{a}}}
\def\vb{{\bm{b}}}
\def\vc{{\bm{c}}}

\def\vp{{\bm{p}}}

\def\vu{{\bm{u}}}
\def\vv{{\bm{v}}}

\def\vx{{\bm{x}}}

\DeclareMathAlphabet{\mathsfit}{\encodingdefault}{\sfdefault}{m}{sl}
\SetMathAlphabet{\mathsfit}{bold}{\encodingdefault}{\sfdefault}{bx}{n}

\def\sR{{\mathbb{R}}}

\newcommand{\R}{\mathbb{R}}

\usepackage{algpseudocode,algorithm,algorithmicx} 

\usepackage[pagebackref=true,breaklinks=true,letterpaper=true,colorlinks,bookmarks=false]{hyperref}

\cvprfinalcopy 

\def\cvprPaperID{
10644} 

\ifcvprfinal\fi
\begin{document}

\title{Temporal Wasserstein non-negative matrix factorization for non-rigid motion segmentation and spatiotemporal deconvolution}
\author{Erdem Varol$^{1,3,4}$,~~Amin Nejatbakhsh$^{2,3,4}$,~~Conor McGrory$^{3,4}$ \\
Department of Statistics$^1$, Department of Neurobiology and Behavior$^2$\\
Center for Theoretical Neuroscience$^3$, Zuckerman Institute$^4$\\
Columbia University\\
}

\maketitle

\begin{abstract}
Motion segmentation for natural images commonly relies on dense optic flow to yield point trajectories which can be grouped into clusters through various means including spectral clustering or minimum cost multicuts. However, in biological imaging scenarios, such as fluorescence microscopy or calcium imaging, where the signal to noise ratio is compromised and intensity fluctuations occur, optical flow may be difficult to approximate. To this end, we propose an alternative paradigm for motion segmentation based on optimal transport which models the video frames as time-varying mass represented as histograms. Thus, we cast motion segmentation as a temporal non-linear matrix factorization problem with Wasserstein metric loss. The dictionary elements of this factorization yield segmentation of motion into coherent objects while the loading coefficients allow for time-varying intensity signal of the moving objects to be captured. We demonstrate the use of the proposed paradigm on a simulated multielectrode drift scenario, as well as calcium indicating fluorescence microscopy videos of the nematode \textit{Caenorhabditis elegans} (\textit{C. elegans}). The latter application has the added utility of extracting neural activity of the animal in freely conducted behavior.
\end{abstract}
\begin{figure}[!htb]
    \centering
    \includegraphics[width=0.5\textwidth]{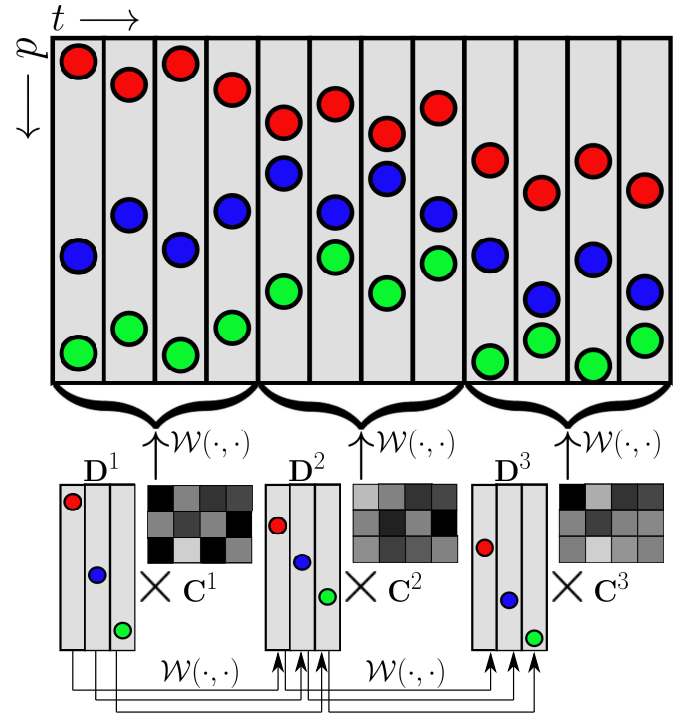}
    \caption{Schematic of the proposed spatio-temporal matrix factorization model. Pixels of correlated movement are locally modelled as being c }
    \label{fig:schematic}
\end{figure}

\begin{figure*}[!htb]
    \centering
    \includegraphics[width=1\linewidth]{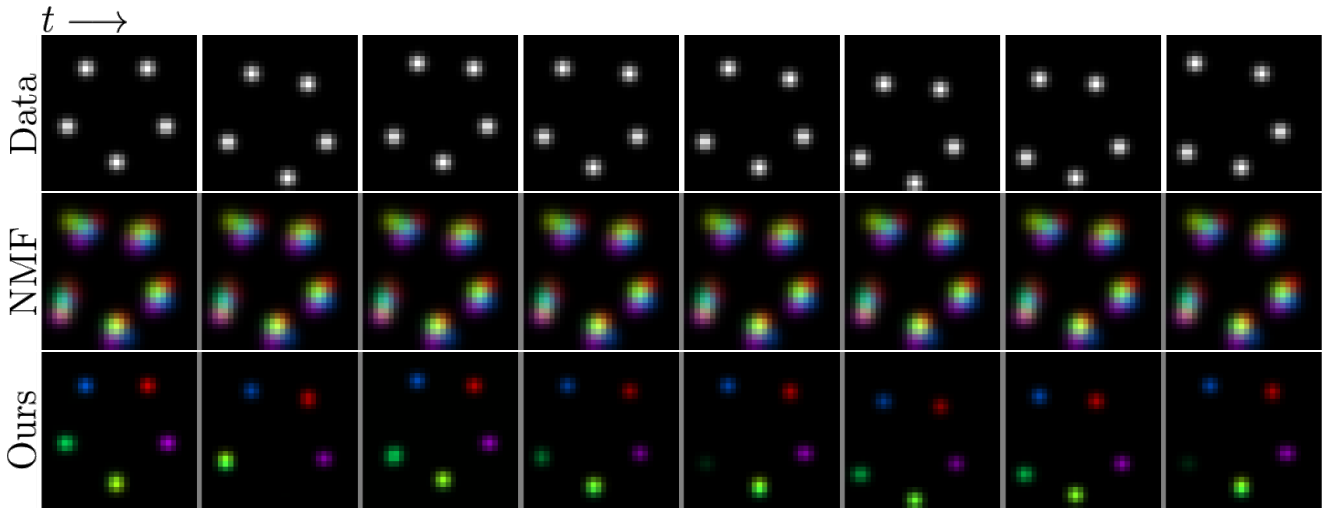}
    \caption{Simulated results of two dimensionmal random walk of five particles \textbf{Top:} Simulated particle data, \textbf{Middle:} Spatio-temporal footprints of NMF components colored by component, \textbf{Bottom:} Spatio-temporal footprints of TWNMF components, colored by component }
    \label{fig:2d_simulation}
\end{figure*}

\section{Introduction}
In the past five years, new imaging techniques\cite{prevedel2014simultaneous} have made it possible to record 3D volumetric data from whole animals with high speed and single-neuron resolution. When used with calcium indicators, these techniques can be used to observe large-scale patterns of neural activity with high spatial and temporal precision. Extracting traces of neural activity from this data is a difficult problem that has been addressed using a variety of approaches. Recently, non-negative matrix factorization (NMF) methods\cite{zhou2018efficient} have been proposed for solving this problem. These methods extract neural activity from imaging data by building a low-rank representation of the data, expressed as a product of two matrices that encode the spatial footprints of neurons and their temporal firing activity respectively. This general approach has proven to be very effective for denoising and demixing calcium signals in a variety of applications. However, one shortcoming of traditional NMF methods is their assumption that the spatial footprint of neurons is invariant across time. In whole-animal recordings of behaving \textit{C. elegans}, this assumption is unrealistic due to the non-linear movement of the worm, causing NMF methods to perform poorly at extracting calcium signals. Tracking methods that explicitly model neuron motion also fail at this task, because they require high-fidelity detection of neurons, which is often not realizable in biological signal-to-noise regimes. Instead, to solve this problem, we propose a novel method called \textit{Temporal Wasserstein Non-negative Matrix Factorization} (TWNMF), a nonlinear extension of traditional NMF methods that allows the spatial footprint of neurons to evolve in time. We describe how this approach can be used to track and extract calcium traces from moving neurons and demonstrate its effectiveness on imaging data of \textit{C. elegans}. We also show how this technique can be used for the slightly related problem of extracting voltage traces from multielectrode array data in the presence of motion drift. 

\subsection{Paper organization}
In section~\ref{sec:related}, we briefly revised the prior work on motion segmentation and optimal transport. In section~\ref{sec:method}, we pedagogically build up our formulation using the classical non-negative matrix factorization and its variants using the Wasserstein metric. In section~\ref{sec:results}, we demonstrate three important applications of our method: multielectrode array signal demixing under motion drift~(\ref{sec:mea_drift}), calcium signal recovery in moving \textit{C.elegans} worms~(\ref{sec:worm_traces}), and motion segmentation of freely moving whole worm body~(\ref{sec:worm_motion}). In section~\ref{sec:conclusion}, we discuss extensions and computational bottlenecks of the proposed method.

\section{Related work and contributions}\label{sec:related}

\subsection{Motion Segmentation}

Motion segmentation is the problem of automatically identifying independently moving objects in a video.\cite{zappella2008motion} Early methods such as \cite{tomasi1992shape} used factorization approaches to accomplish this, under the assumption that the objects in question were undergoing rigid motion. More recent approaches solve this problem by estimating \textit{optical flow} -- the inferred "motion" of pixels between successive images. \cite{anthwal2019overview} Optical flow can be estimated in a variety of ways, including using deep neural networks \cite{ilg2017flownet}. Once the optical flow between each successive pair of images has been estimated, objects can be identified by either clustering the long-term point trajectories derived from the flows \cite{ochs2013segmentation} \cite{xie2019object}, modeling the pixels and trajectories as a graph and computing minimum-cost multi-cuts \cite{keuper2015motion}, or using additional features to group the trajectories \cite{bideau2018best} \cite{taylor2015causal} \cite{dave2019towards}. The major problem with applying these approaches to calcium imaging data is that they perform badly when the pixel intensity of the moving objects in the image changes rapidly, which happens in scenarios where calcium indicators are used to measure neural activity.


\subsection{Optimal Transport Methods}

Optimal transport theory \cite{villani2008optimal} has been recently applied to image processing problems, specifically through the use of the Wasserstein distance as a similarity metric between images. Cuturi et al. \cite{cuturi2014fast} use this metric to compute prototypical images from sets of similar images. Rolet et al. \cite{rolet2016fast} and Quian et al. \cite{qian2016non} use it as the loss function for non-negative matrix factorization methods, an approach that we call \textit{Wasserstein non-negative matrix factorization} (WNMF). Our proposed method here is an extension of the WNMF algorithm that incorporates temporal connections and constraints.

\subsection{Contributions}

Our contributions are three-fold:
\begin{itemize}
    \item We cast non-rigid motion segmentation as a first-principles non-linear matrix factorization problem utilizing the geometric structure that the Wasserstein metric provides for data fidelity terms.
    \item The basis elements resulting from the matrix factorization provide spatio-temporal footprints of moving objects while the temporal coefficients capture the temporal appearance intensities effectively modeling motion segmentation and spatio-temporal deconvolution simultaneously.
    \item We apply the proposed technique to demixing simulated voltage data in motion drifting multielectrode scenarios.
    \item We demonstrate the ability to capture neural activity in the moving worm \textit{C. elegans} in a completely unsupervised way for the first time.
\end{itemize}

\section{Method}\label{sec:method}
First we introduce notation. Let $\textbf{X} \in \mathbb{R}^{D \times T}$ be a non-negative matrix containing the imaging data, with $T$ frames of $d$ pixels or voxels each.  
The goal of non-negative matrix factorization (NMF) is to represent $\textbf{X}$ as a product of two non-negative matrices $\textbf{D} \in \mathbb{R}^{D \times K}$ and $\textbf{C} \in \mathbb{R}^{K \times T}$, where $K \ll d$, effectively compressing $\textbf{X}$ to a lower-dimensional latent space. Each column of $\mathbf{D}$ represents the spatial footprint of one of the $K$ latent \textit{components}, and each row of $\mathbf{C}$ represents its activity across time. To compute $\mathbf{D}$ and $\mathbf{C}$ for a given $\mathbf{X}$, we minimize the cost function $E_{NMF}(\mathbf{D}, \mathbf{C}) = \|\mathbf{X} - \mathbf{DC}\|_{F}^2$, subject to the constraint that $\mathbf{D}$ and $\mathbf{C}$ are non-negative. This can be done efficiently using a simple iterative procedure~\cite{lee2001algorithms}. If the signal-to-noise ratio in our data is sufficiently large, the components we recover correspond to spatial regions with correlated temporal activity, with $\mathbf{D}$ encoding their spatial footprints and $\mathbf{C}$ encoding their relative intensities over time. However, one of the assumptions in NMF is that the spatial footprint of time-varying activity is fixed in time and does not model the possibility of motion.

\begin{algorithm}[!htb]
\caption{[$\boldsymbol{\alpha}$,$\boldsymbol{\beta}$,$\boldsymbol{\pi}$]= SINKHORN($\va,\vb,\rmM,\gamma$)}\label{alg:sinkhorn}
\begin{algorithmic}
\State{\textbf{Input: } $\va,\vb\in \mathbb{R}^d$ such that $\va\geq 0,\vb\geq 0$, $\va^T\boldsymbol{1} = 1, \vb^T\boldsymbol{1} = 1$ (input histograms), $\rmM \in \mathbb{R}^{d\times d}$ (ground cost), $\gamma \geq 0$ (entropic regularization)}
\State{\textbf{Set:}} $\mathbf{K} = e^{-\gamma \rmM}$, $\boldsymbol{u} = \boldsymbol{1}_d$,$\boldsymbol{v} = \boldsymbol{1}_d$  
\While{not converged}
\State $\boldsymbol{u} \leftarrow \va / \rmK \boldsymbol{v}$ (elementwise)
\State $\boldsymbol{v} \leftarrow \vb / \rmK \boldsymbol{u}$ (elementwise)
\EndWhile
\Return{Dual variables $\boldsymbol{\alpha} = -\gamma \log \vu$, $\boldsymbol{\beta}   = -\gamma \log \vv$, transportation plan $\boldsymbol{\pi} = diag(\vu) \rmK \text{diag}(\vv)$}
\end{algorithmic}
\end{algorithm}

\begin{algorithm}[!htb]
\caption{Temporal Wasserstein Non-negative Matrix Factorization (TWNMF)}\label{alg:twnmf}
\begin{algorithmic}
\State{\textbf{Input: } Signal: $\rmX \in \sR^{d \times T}$, ground cost: $\rmM \in \sR^{d \times d}$, entropic regularization: $\gamma > 0$, Temporal entropy penalty: $\lambda_T \geq 0$, Spatial entropy penalty: $\lambda_D$,Step size: $\nu\geq 0$}
\State{\textbf{Require:} Data normalization: $\rmX_t^T\boldsymbol{1} = 1$, $\rmX \geq 0$}
\State{\textbf{Initialization:}$\rmD_{\ell}\leftarrow \text{Unif}[0,1]^{d\times K}$ (column sum normalized), $\rmC_{\ell}\leftarrow \text{Unif}[0,1]^{K\times |T_\ell|}$ (row sum normalized such that $\rmD_\ell\vc_t$ is a histogram)}
\While{not converged}
\For{$\ell=1,\ldots,L$}
\For{$t\in T_\ell$}
\State{ $[\boldsymbol{\alpha}^{\ell,t},\boldsymbol{\beta}^{\ell,t},\boldsymbol{\pi}^{\ell,t}]\leftarrow$SINKHORN($\vx_t$,$\rmD_{\ell}\vc_t$,$\rmM$,$\gamma$)}
\State{ $[\boldsymbol{\psi}^{\ell,k},\boldsymbol{\omega}^{\ell,k},\boldsymbol{\rho}^{\ell,t}]\leftarrow$SINKHORN($\rmD_{\ell}^k$,$\rmD_{\ell+1}^k$,$\rmM$,$\gamma$)}
\For{k=1,\ldots,K}
\State{\hspace{-7ex}$\rmD_{\ell}^k \leftarrow [\rmD_{\ell}^k - \nu \nabla_{\rmD_{\ell}^k}\mathcal{L}(\rmD,\rmC,\boldsymbol{\pi},\boldsymbol{\rho},\boldsymbol{\alpha},\boldsymbol{\beta},\boldsymbol{\psi},\boldsymbol{\omega})]_+$}
\State{\hspace{-7ex}$\vc_{t} \leftarrow [\vc_{t} - \nu \nabla_{\vc_t}\mathcal{L}(\rmD,\rmC,\boldsymbol{\pi},\boldsymbol{\rho},\boldsymbol{\alpha},\boldsymbol{\beta},\boldsymbol{\psi},\boldsymbol{\omega})]_+$}
\State{\hspace{-7ex}Column normalize $\rmD_{\ell}^k$}
\State{\hspace{-7ex}Row normalize $\vc_{t}$ such that $\rmD_{\ell}^k\vc_t$ is a histogram}
\EndFor
\EndFor
\EndFor
\EndWhile\\
\Return{Spatiotemporal footprints: $\lbrace \rmD_{\ell}\rbrace_{\ell=1}^L$, spatiotemporal loading coefficients: $\lbrace\rmC_{\ell}\rbrace_{\ell=1}^L$, Dictionary-to-data segmentation push-forwards: $\lbrace \boldsymbol{\rho}^{\ell,t} \rbrace$}
\end{algorithmic}
\end{algorithm}

\begin{figure*}[!htb]
    \centering
    \includegraphics[width=1\linewidth]{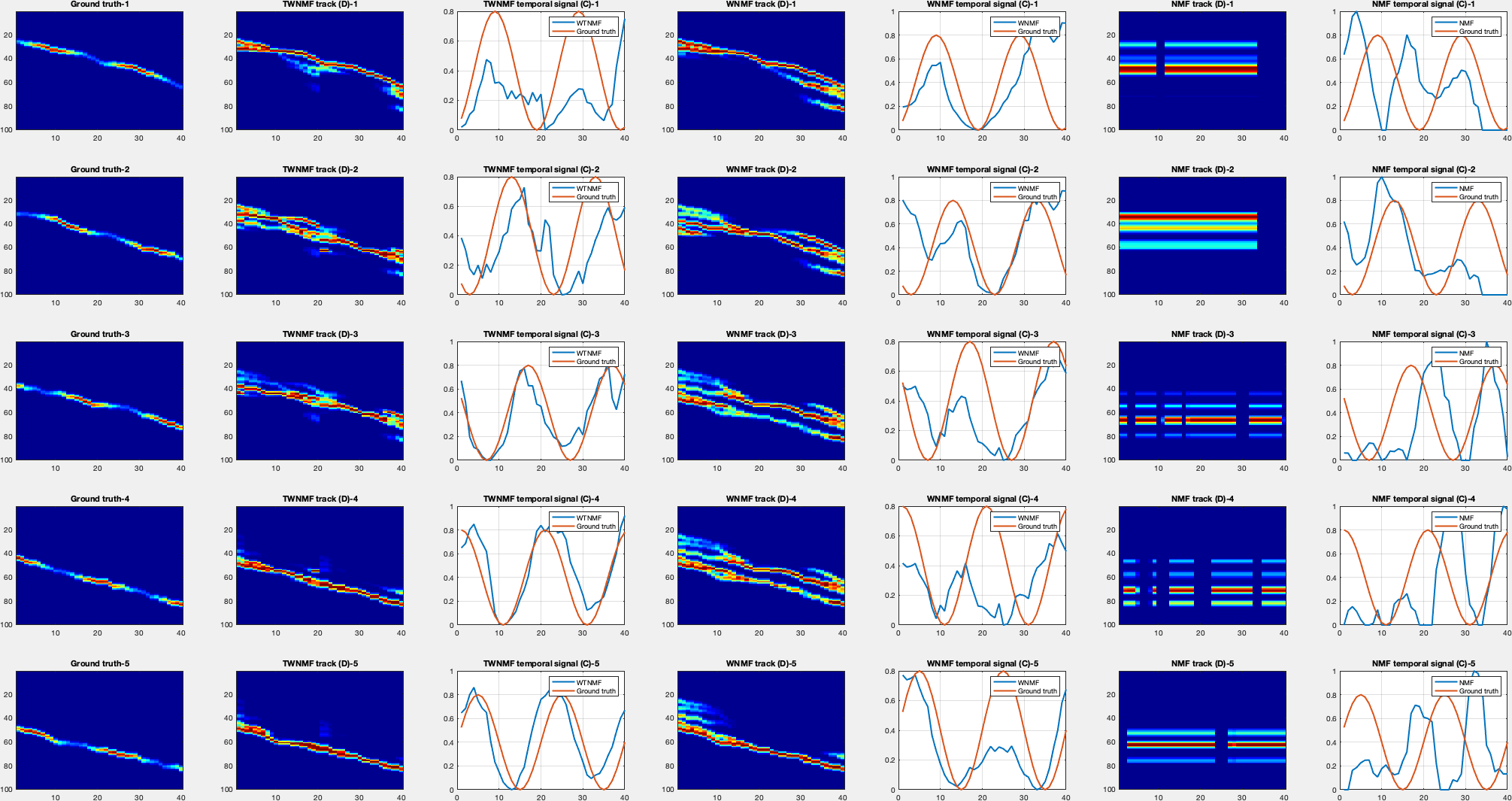}
    \caption{One dimensional data simulating multielectrode array motion drift, x-axis in subplots denotes the electrode positions, while y-axis denotes the voltage captured across time. Rows correspond to spatio-temporal footprints of five different neurons measured by the electrode array. \textbf{1st column:} The ground truth spatio-temporal footprints of five neurons, \textbf{2nd column:} Spatio-temporal footprints of TWNMF components, \textbf{3rd column:} Temporal components of WNMF (blue) compared with actual voltage signal (red), \textbf{4th column} Spatio-temporal footprints of WNMF\cite{rolet2016fast} components, \textbf{5th column:} Temporal components of WNMF (blue) compared with actual voltage signal (red), \textbf{6th column:}Spatio-temporal footprints of NMF\cite{lee2001algorithms} components, \textbf{7th column:} Temporal components of NMF (blue) compared with actual voltage signal (red)}
    \label{fig:drift_simulation}
\end{figure*}
\begin{figure}[!htb]
    \centering
    \includegraphics[width=1\linewidth]{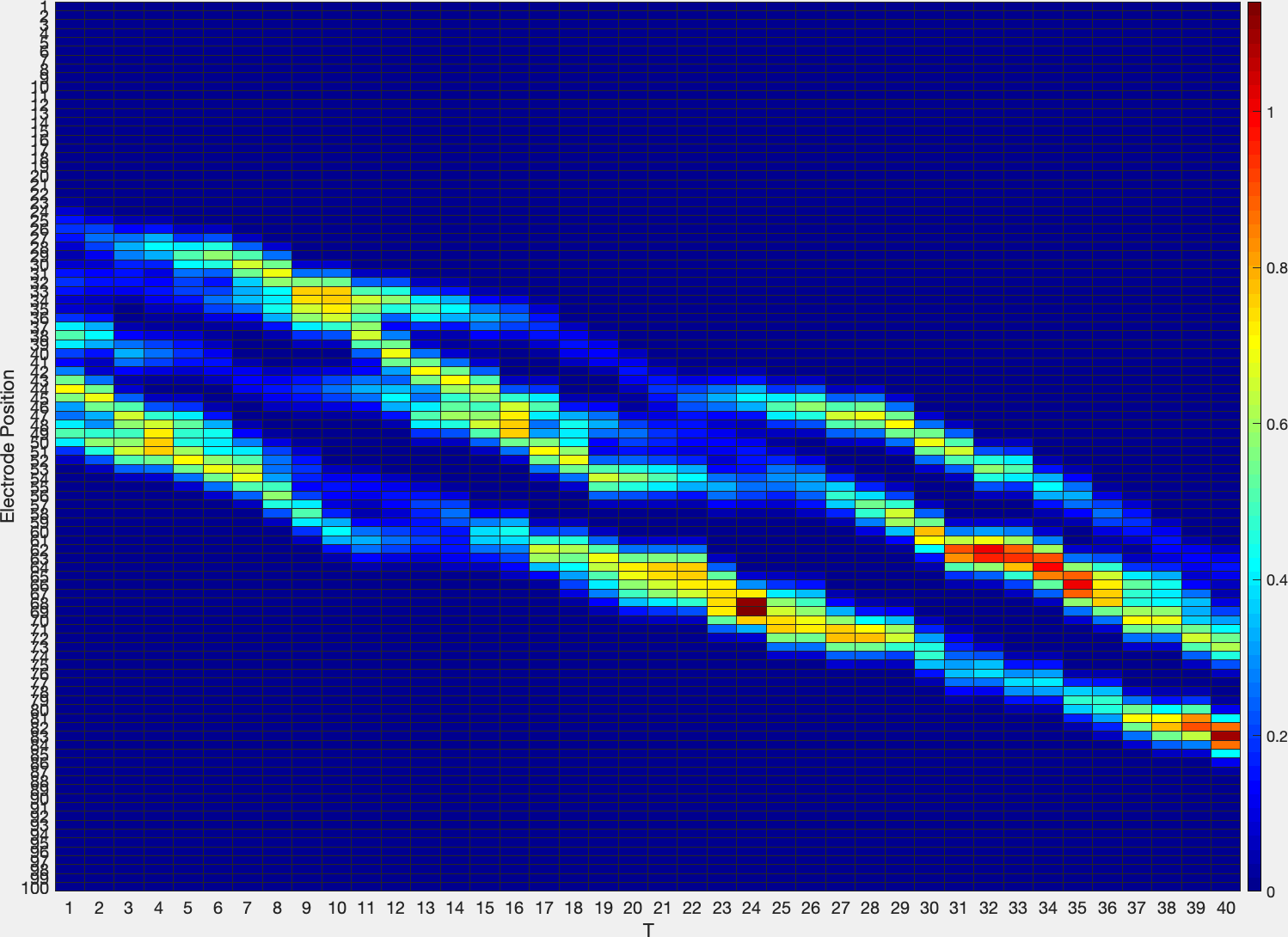}
    \caption{Simulated multielectrode array voltage data.}
    \label{fig:electrode_data}
\end{figure}
\subsection{Wasserstein distance} Wasserstein distances have gained popularity in recent years as a way to measure the similarity between images. These distances work by solving different forms of the \textit{optimal transport problem}, a well-studied problem that, in its most general form, asks how to efficiently transport mass from one spatial distribution to another, given an underlying cost metric\cite{villani2008optimal}. To compute the OT distance between two images, we treat pixel intensity as mass and use the euclidean distance between pixels as the transportation cost. Computing this exactly is prohibitively expensive for most image-processing applications, but fast and accurate approximations \cite{cuturi2013sinkhorn} exist. 

For two input histograms $\va,\vb \in \R^d$, the Wasserstein metric $W_1$ can be expressed as the solution to a constrained linear program:
\begin{align}\label{eq:wasserstein}
    &W_1(\va,\vb) = \underset{\rmP}{\min}~\sum_{i,j=1}^{d} m_{i,j}P_{i,j}  \nonumber \\
    &\text{subject to}~\sum_{j=1}^d P_{i,j} = \va,~~\sum_{i=1}^d P_{i,j} = \vb 
\end{align}
where $m_{i,j}$ is the ground cost of moving a unit of mass from histogram bin $i$ to histogram bin $j$. The ground cost is typically set to be the euclidean distance of the pixel locations i.e. $m_{i,j} = \|\vp_i -\vp_j\|_2$ where $\vp_i$ is the coordinates of the ith pixel. Solving this linear program in general requires $O(d^3)$ operations and furthermore may not yield a unique optima since it is possible for entire facets of the feasible region polytope (i.e. $\lbrace \mathcal{P}: \rmP \in\mathbb{R}^{d\times d}~~|~~ \rmP\boldsymbol{1} = \va, \rmP^T \boldsymbol{1} = \vb \rbrace$ to yield a minimum cost solution.

Cuturi~\cite{cuturi2013sinkhorn} has proposed a relaxation of~\eqref{eq:wasserstein} by introducing an entropic regularizer:
\begin{align}
    &W_1^\gamma(\va,\vb) = \underset{\rmP}{\min}~\sum_{i,j=1}^{d} m_{i,j}P_{i,j} + \gamma (P_{i,j}\log P_{i,j} -1) \nonumber \\
    &\text{subject to}~\sum_{j=1}^d P_{i,j} = \va,~~\sum_{i=1}^d P_{i,j} = \vb 
\end{align}
which is strongly convex and can be optimized using fixed point Sinkhorn algorithm\cite{sinkhorn1967concerning} iterations in $O(d\log d)$ operations. The outline of the Sinkhorn algorithm is provided in algorithm~\ref{alg:sinkhorn}.

\begin{figure*}[!htb]
    \centering
    \includegraphics[width=1\linewidth]{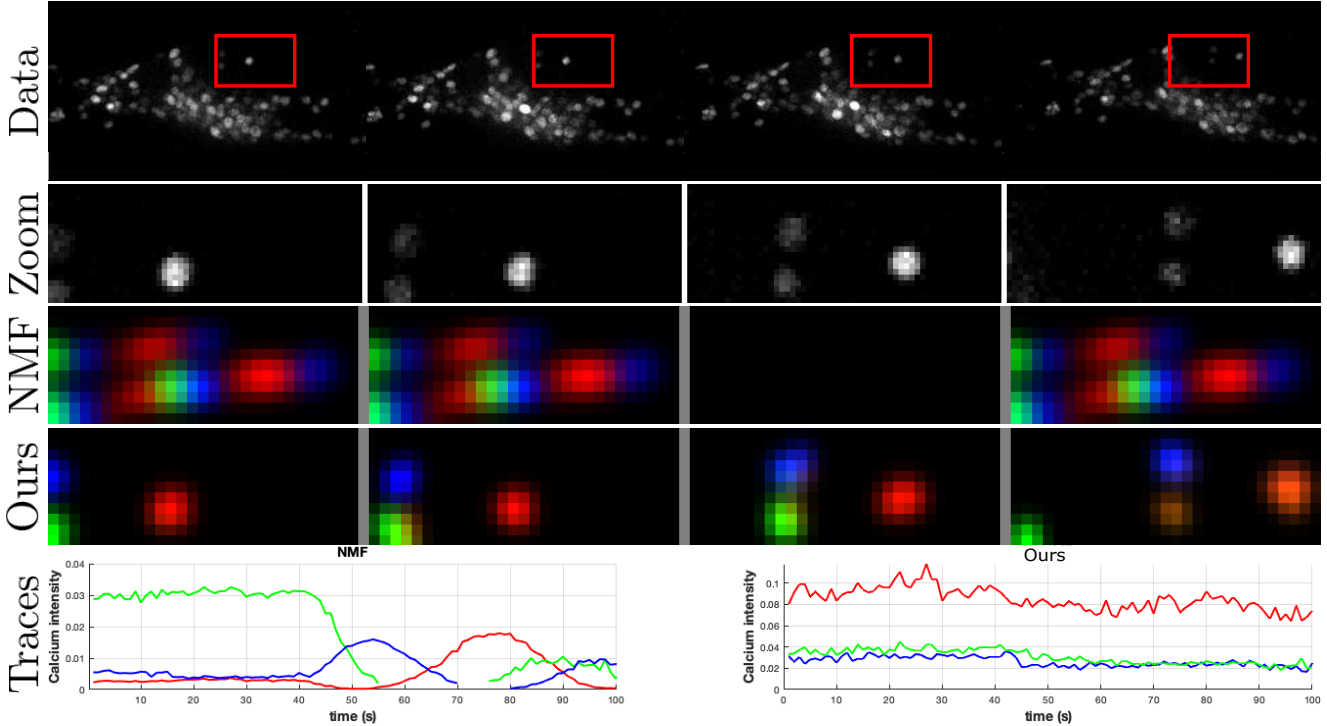}
    \caption{\textbf{Top:} Moving \textit{C. elegans} fluorescence microscopy video, \textbf{2nd row:} Zoomed region, \textbf{3rd row:} Spatio-temporal footprints of NMF components colored by component, \textbf{4th row:} Spatio-temporal footprints of W-NMF components, colored by component, \textbf{Bottom: } Calcium traces extracted from NMF (left) and W-NMF (right) components.}
    \label{fig:worm_video}
\end{figure*}

\subsection{Wasserstein non-negative matrix factorization}
While the euclidean loss takes into account the magnitude difference two vectors, it does not take into account the underlying manifold that the bases of the vectors reside in. On the other hand, the wasserstein metric not only takes into account the differences in the loadings of the basis vectors but also the distances of the basis vectors itself. To illustrate this, let $\va = [1,0,0]$,$\vb =[0,1,0]$ and $\vc=[0,0,1]$ be three vectors in $\mathbb{R}^3$. Under the euclidean metric $d_E$, $d_E(\va,\vb) = d_e(\va,\vc)$. However, if we endow this metric space with the ground cost matrix $\rmM=\bigg[\begin{matrix}0 & 1 & 2\\1 & 0 & 1\\2 & 1 & 0 \end{matrix}\bigg]$, then it is easy to see that $W_1(\va,\vb) = 1$ and $W_1(\va,\vc) = 2$. Thereby, if there were to be a flow of movement from $\va$ to $\vb$ or $\vc$, the motion $\va \rightarrow \vb$ would be the most likely one.

Thus, the transportation plan $\rmP$ that is computed to obtain the Wasserstein metric inherently resembles a flow field of mass from source histogram to the target histogram, making it an attractive choice for modeling motion and mass transfer.

Thus using the Wasserstein metric instead of the euclidean distance in the loss function non-negative matrix factorization allows for wiggle room for the spatial footprints to capture motion varying signal captured in the data matrix $\rmX$.  Intuitively, by using the Euclidean loss, NMF methods assume that object shapes in the data can be explained as a set of static spatial footprints perturbed by independent pixel noise. For WNMF methods, this perturbation is replaced by 'motion noise' -- a more reasonable assumption for the case where the objects are moving. The first use of the entropically regularized Wasserstein metric in non-negative matrix factorization was introduced by Rolet et al.\cite{rolet2016fast} and expanded by Schmitz et al. \cite{schmitz2018wasserstein}. The new cost function minimized by WNMF is $E_{WNMF}(\mathbf{Y}, \mathbf{C}) = \sum_{t = 1}^T W^\gamma(\rmX_t, \mathbf{D}\vc_t)$, where $W^\gamma$ is the smoothed Wasserstein distance.

While the advantage of WNMF is that it can capture spatial footprints of moving objects, the fundamental limitation of it is that, if the motion variability is too high due to very long time courses, the uncertainty in the flow fields from a static dictionary to arbitrarily distant (in a Wasserstein sense) data entries, yields poor reconstruction and thus low data fidelity.

\subsection{Temporal Wasserstein non-negative matrix factorization}
To alleviate the problem of modeling large motion variability using a single dictionary, we propose to use a temporally constrained piecewise Wasserstein non-negative matrix factorization. In our formulation, we break the data matrix $\rmX$ into blocks of length $L$ and model that each of these blocks is factored as a product of a dictionary $\rmD_\ell$ and a coefficient matrix $\rmC_\ell$. To model for motion within these time blocks, we utilize the entropic Wasserstein metric as the data fidelity loss. To yield temporally coherent motion tracks, we further constrain the corresponding terms in the sequential dictionary blocks to be close in the Wasserstein metric space by adding the terms $W(\rmD_\ell^k,\rmD_{\ell+1}^k)$. Furthermore, we penalize the negative entropy, $H(\rmD)$, of the dictionary elements to yield spatially compact regions. Lastly, we penalize the entropy of the loading coefficients to yield spatially smooth signals $H(\rmC)$. This formulation yields the objective function below:

\begin{align}\label{eq:twnmf}
    &\underset{\rmD,\rmC}{\min}\sum_{\ell=1}^L \sum_{t \in T_\ell} W^\gamma(\vx_t,\rmD_\ell \vc_t)  + \lambda_T \sum_{\ell=1}^L\sum_{k=1}^K W^\gamma(\rmD_\ell^k,\rmD_{\ell+1}^k) \nonumber \\
    &- \lambda_D H(\rmD) - \lambda_C H(\rmC)    
\end{align}
We term this model, temporal Wasserstein non-negative matrix factorization (TWNMF). As in NMF, this objective is non-convex but convex in $\rmD$ and $\rmC$, separately. To optimize the objective, we form the Lagrangian system that incorporates the dual variables $\boldsymbol{\alpha,\beta,\psi,\omega}$ to account for the Wasserstein metric constraints:
\begin{align}\label{eq:lagrangian}
&\mathcal{L}(\rmD,\rmC,\boldsymbol{\pi},\boldsymbol{\rho},\boldsymbol{\alpha},\boldsymbol{\beta},\boldsymbol{\psi},\boldsymbol{\omega})\nonumber\\
&\sum_{\ell=1}^L \sum_{t \in T_\ell} \sum_{i=1}^d\sum_{j=1}^d \pi_{i,j}^{\ell,t}m_{i,j} +\gamma(\pi_{i,j}^{\ell,t}\log\pi_{i,j}^{\ell,t} - 1)\nonumber\\
    &+ \lambda_T \sum_{\ell=1}^L\sum_{k=1}^K\sum_{i=1}^d\sum_{j=1}^d \rho_{i,j}^{\ell,k}m_{i,j}+\gamma(\rho_{i,j}^{\ell,k}\log\rho_{i,j}^{\ell,k} - 1) \nonumber\\
    &+ \lambda_D \sum_{\ell=1}^L \sum_{k=1}^K \sum_{i=1}^d (D_{\ell,i}^k\log D_{\ell,i}^k -1)\nonumber\\
    &+ \lambda_C \sum_{\ell=1}^L \sum_{k=1}^K \sum_{t\in T_\ell} (C_{\ell,t}^k\log C_{\ell,t}^k -1)\nonumber\\
    &+ {\boldsymbol{\alpha}^{\ell,t}}^T(\boldsymbol{\pi}^{\ell,t}\boldsymbol{1} - \vx_t) + {\boldsymbol{\beta}^{\ell,t}}^T({\boldsymbol{\pi}^{\ell,t}}^T\boldsymbol{1} - \rmD_{\ell}\vc_t) \nonumber\\
    &+{\boldsymbol{\psi}^{\ell,k}}^T({\boldsymbol{\rho}^{\ell,k}}\boldsymbol{1} - \rmD_{\ell}^k)+ {\boldsymbol{\omega}^{\ell,k}}^T({\boldsymbol{\rho}^{\ell,k}}^T\boldsymbol{1} - \rmD_{\ell+1}^k)
\end{align}
The optimization routine for \eqref{eq:twnmf} is outlined in algorithm~\ref{alg:twnmf}. For each time block, we optimize for the dual variables $\boldsymbol{\alpha,\beta}$ using the Sinkhorn 
iteration using the subroutine~\ref{alg:sinkhorn}. 
Furthermore, the Sinkhorn iterates are used to optimize the temporal glueing dual variables $\boldsymbol{\psi,\omega}$. Once the dual variables are updated, we perform projectced gradient descent on the dictionary elements $\rmD$ and $\rmC$, respecting non-negativity. To obtain spatial tracks from $\rmD^{\ell,k}$, we use the transportation plan $\boldsymbol{\rho}_\ell$ to obtain the push-forward flow field onto the data matrix using $\rmT = \boldsymbol{\rho}^{\ell,t}\rmD^{\ell,k}$.

\section{Results}\label{sec:results}
To demonstrate the utility of our method in performing Spatio-temporal deconvolution, we experimented with three different applications. First, we attempted to tackle the motion drift problem in multielectrode array voltage deconvolution. Then we showed that TWNMF can be used to extract calcium signal in freely moving C. elegans videos in full spatial resolution, constraining our analysis to a subregion of voxels such that the computational cost of performing optimal transport is manageable. Lastly, we attempted to motion segment a full-body video of \textit{C.elegans} after it has gone dimensionality reduction using the matching pursuit algorithm.

\subsection{Multielectrode array motion drift}\label{sec:mea_drift}

To test the effectiveness of TWNMF for recovering neural voltage traces from multielectrode array recordings\cite{obien2015revealing,spira2013multi} in the case of motion drift, we created data simulating 5 neurons recorded by a drifting multielectrode array with 100 electrodes. We model five neurons drifting in a one-dimensional random walk while exhibiting a random phase changed sinusoidal voltage pattern. We then ran the TWNMF, WNMF\cite{rolet2016fast}, and NMF\cite{zhou2018efficient} algorithms on the time series of electrode values, and compared the traces extracted by each algorithm to the true voltage traces (Figure \ref{fig:drift_simulation}). In this simulation, TWNMF recovered the original traces with the highest accuracy. The traces extracted by WNMF were reasonably accurate for the middle time points but differed significantly from the true signals at time points close to the beginning and end of the series. This is most likely because the WNMF method models all object motion as noise, and therefore cannot extract traces from cells that move significantly from their original positions. The traces extracted by NMF sometimes captured part of the true signal but were inaccurate for most time points. This is because NMF does not model cell motion at all, and attempts to construct a static representation for all cells.


\begin{table}[!htb]
    \centering
    \begin{tabular}{|l|l|l|}
    
    \hline
        \textbf{Method} & \textbf{Spatial accuracy} & \textbf{Temporal accuracy}\\
        \hline
        \hline
         TWNMF & \textbf{0.81 $\pm$ 0.07} & \textbf{0.62 $\pm$ 0.33}\\
         \hline
         WNMF\cite{rolet2016fast} & 0.72 $\pm$ 0.13 & 0.39 $\pm$ 0.21\\
         \hline
         NMF\cite{zhou2018efficient} & 0.27 $\pm$ 0.37 & 0.05 $\pm$ 0.30\\
         \hline
         
    \end{tabular}
    \caption{Quantitative evaluation of the proposed method, WNMF, and NMF on the simultaneous motion segmentation and voltage demixing in the simulated multielectrode drift data. Proposed TWNMF recovers the spatial footprints of drifting neuron signal with high accuracy while at the same time, the temporal voltage signal is accurately deconvolved.}
    \label{tab:electrode_data}
\end{table}

\begin{figure*}[!htb]
    \centering
    \includegraphics[width=1\linewidth]{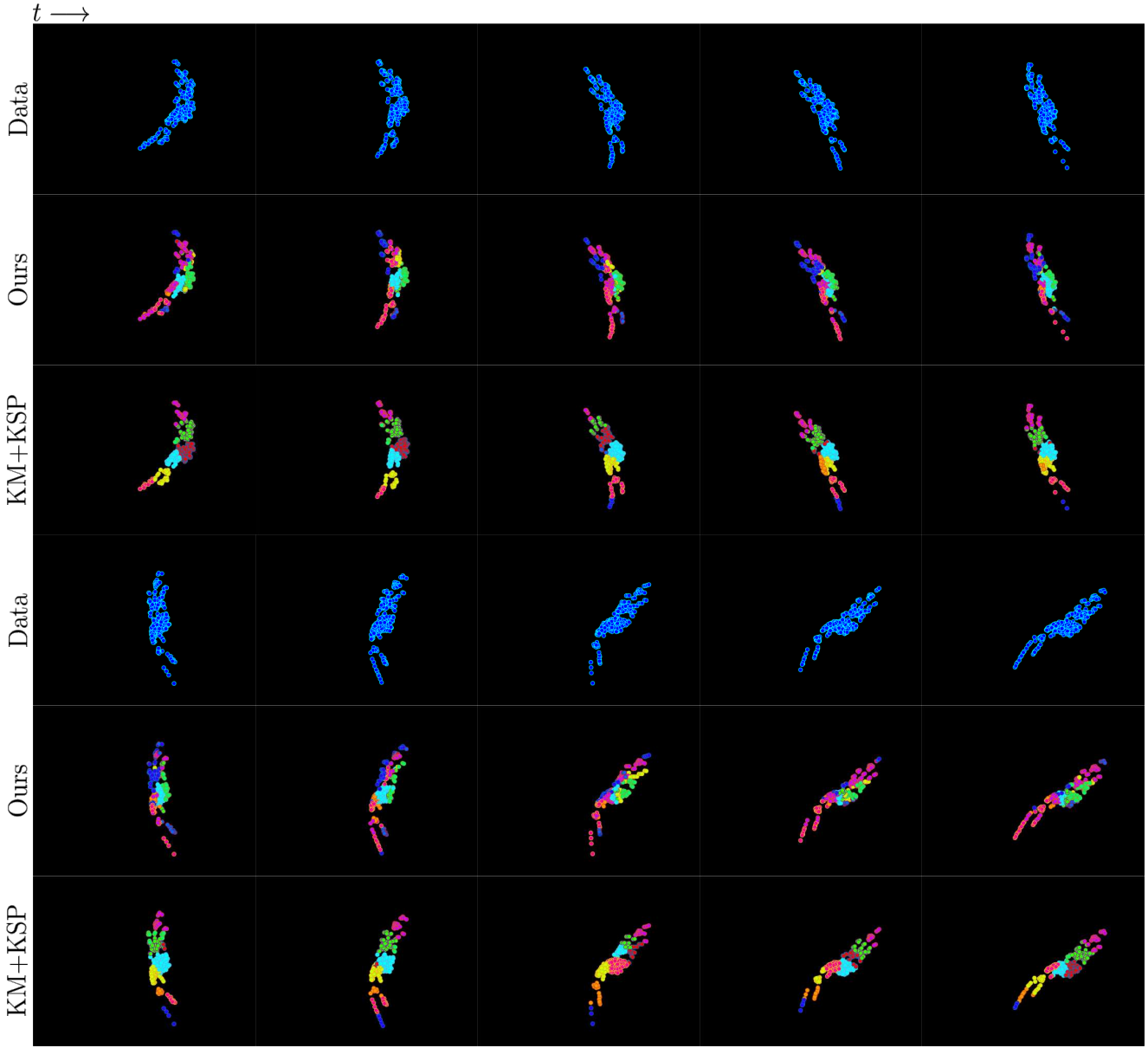}
    \caption{Motion segmentation of moving \textit{c. elegans} worm using proposed method and K-means with K-shortest paths algorithm (KM+KSP). Top: The raw point cloud of matching pursuit hits of C. elegans neurons. Middle: Motion segmentation components using proposed TWNMF algorithm. Bottom: Motion segmentation components using K-means stiched together with K-shortest paths\cite{berclaz2011multiple}.}
    \label{fig:worm_mp}
\end{figure*}

\subsection{Extracting calcium traces from C. elegans videos} \label{sec:worm_traces}

Next, we used TWNMF to try to extract calcium traces from real \textit{C. elegans} imaging data (Figure \ref{fig:worm_video}). For computational efficiency purposes, the sample video we used was a sum-projection of a zoomed-in section of the original 3D video. For comparison, we also extracted traces using NMF. The spatial components extracted by TWNMF correctly identified and tracked the cells, and the calcium traces extracted by the algorithm also appear reasonable. The components extracted by NMF, however, are much more distributed spatially, and their corresponding calcium traces do not look realistic.

\subsection{Motion segmentation in C. elegans videos}\label{sec:worm_motion}

To see if TWNMF could track groups of cells in \textit{C. elegans} imaging data, we ran it on a 3D, full-worm fluorescent microscopy video and plotted the time series of components it extracted (Figure \ref{fig:worm_mp}). The video was taken from a freely behaving, transgenic strain of \textit{C. elegans} called NeuroPAL \cite{yemini2019neuropal}. The video was imaged at a temporal resolution of 5Hz and spatial resolution of 0.3 $\mu m$ $\times$ 0.3 $\mu m$ $\times$ 1 $\mu m$  (x,y,z) at 900 $\times$ 600 $\times$ 30 voxels.  To make TWNMF computationally tractable on the data, each frame was represented as a weighted combination of 500 Gaussian functions with identical covariances, whose means and weights were computed using a greedy matching-pursuit (MP) algorithm \cite{elad2010sparse}. For comparison, a more traditional motion segmentation approach was also applied to the data. In this method, the Gaussian components for each frame were first clustered using the k-means algorithm \cite{lloyd1982least} and then connected across frames using a k-shortest-paths algorithm \cite{berclaz2011multiple} to produce the tracked groups of cells.

The motion segmentation results of TWNMF demonstrate the ability to accurately track consistent parts of the \textit{C.elegans} worm under large motion between frames. In comparison, performing K-means clustering which is followed by K-shortest paths routing to stitch together segments yields a motion segmentation that is highly inconsistent over time, where clusters highly drift from one part of the worm to another.

\section{Conclusion}\label{sec:conclusion}

In this paper, we advocated for a new paradigm for motion segmentation using optimal transport theory. Using the Wasserstein metric to encourage data fidelity and temporal continuity enables the formulation of motion segmentation of non-rigid moving bodies as a non-linear matrix factorization problem. This contrasts with the latest state of the art techniques for motion segmentation which heavily rely on optical flow. Our formulation can be cast a simple piecewise convex factorization model which has the advantage of being interpretable in its parts and allows for further modeling extensions. A further advantage of the proposed model is that unlike state of the art optical flow based motion segmentation models that work on natural images\cite{keuper2015motion}, the temporal coefficients of the model allow for capturing temporal variability in appearance and brightness intensity of the objects tracked.

Thus we demonstrate that the proposed method works well in biological imaging scenarios where the intensities of the moving objects fluctuate in time. We demonstrated its effectiveness for extracting voltage traces from multielectrode array data in the case of motion drift, as well as tracking and extracting activity traces from neurons in \textit{C. elegans} calcium imaging data. One drawback to the method we have proposed is its computational cost -- storing the optimal transport plan in memory is prohibitively expensive for high-resolution images. Just like in many of the optimal flow-based methods, we avoid this problem by using dimensionality reduction techniques to reduce the memory cost of computing optimal transport. Improving these dimensionality reduction methods will be an important part of further work.
\subsection*{Acknowledgements}
The authors would like to acknowledge the sources of funding: NSF NeuroNex Award DBI-1707398, and The Gatsby Charitable Foundation.

{\small
\bibliographystyle{ieee}
\bibliography{refs}
}

\end{document}